\documentclass[twocolumn,prb,aps,showpacs]{revtex4}

\usepackage{psfrag,graphicx}
\usepackage{dcolumn}
\usepackage{amsmath,amssymb}
\usepackage{bm}
\usepackage{pstricks}
\usepackage{hyperref}
\usepackage{epsfig}

\def\<{\left<}
\def\>{\right>}
\def\ket|#1>{\left|#1\right>}
\def\bra<#1|{\left<#1\right|}
\def\elem<#1|#2|#3>{\left<#1\right|#2\left|#3\right>}
\def\({\left(}
\def\){\right)}

\begin{document}

\title{Density matrix renormalization on random graphs and the quantum
spin-glass transition}

\author{J. Rodr\'{\i}guez-Laguna}
\affiliation{International School for Advanced Studies (SISSA), Via
Beirut 2-4, I-34014 Trieste, Italy}

\date{\today}

\begin{abstract}
The density matrix renormalization group (DMRG) has been extended to
study quantum phase transitions on random graphs of fixed
connectivity. As a relevant example, we have analysed the random Ising
model in a transverse field. If the couplings are random, the number
of retained states remains reasonably low even for large sizes. The
resulting quantum spin-glass transition has been traced down for a few
disorder realizations, through the careful measurement of selected
observables: spatial correlations, entanglement entropy, energy gap
and spin-glass susceptibility, among others.
\end{abstract}


\maketitle

\section{Introduction} \label{introd} 

Quantum phase transitions constitute a very active topic in
theoretical physics \cite{Sachdev:book}, to which disorder and
frustration of a quantum spin glass add considerable intricacies
\cite{Rieger:chapter}. Although quantum effects were usually
considered to be negligible in spin glasses at finite temperatures,
diverse experiments \cite{Wu_PRL93,Maclaughlin_PRL01,Chen_PRB05} have
proved them to be highly relevant, leading to the first experimental
realization of {\em quantum annealing} \cite{Brooke_Sci99}. One of the
most widely employed theoretical approaches to the quantum spin-glass
transition (QSGT) is the random couplings Ising model in a transverse
field (RITF) \cite{Chakrabarti_Dutta_Sen:book}. Within the analytical
approach, a Griffiths phase was found in the 1D case via an elegant RG
procedure \cite{Fisher_PRB95,Fisher_PRB98}, while such a structure was
proved to be absent from the infinite range case within the replica
formalism \cite{Bray_JPC80,Goldschmidt_PRL90,Miller_PRL93}. The 1D RG
analysis was extended to study the presence of a certain number of
long-distance links, proving the relevance of the perturbation
\cite{Carpentier_PRE05}. Regarding numerical approaches, quantum
Monte-Carlo (QMC) remains as the most suitable tool at $T>0$. Using
it, a coherent picture was obtained in the 2D and 3D cases, showing
that the Griffiths phase is present in 2D but absent in 3D
\cite{Guo_PRL94,Rieger_PRL94,Rieger_PRB96}.

Classical spin glasses have been thoroughly analysed on Bethe lattices
\cite{Thouless_PRL86}. However, finite Bethe lattices are dominated by
boundary effects and present no frustration. A way out is to study
random graphs with a fixed connectivity which, in the classical case,
allow the application of the {\em cavity approach}
\cite{Mezard_EPJ01}. These random graphs present genuine frustration,
since they have {\em loops}, and have no boundaries. On the other
hand, the average size of the loops grows with the number of sites
\cite{Marinari_JSTAT04}, thus making them resemble, {\em locally}, a
Bethe lattice (see figure \ref{graph:fig}).

The density matrix renormalization group (DMRG)
\cite{White_PRL92,Schollwock_RMP05} is known to be a highly accurate
method to analyze $T=0$ ground state properties of 1D or quasi-1D
quantum many body systems, including tree structures
\cite{Martin_PRB02}. We have extended the method in order to trace the
behaviour of finite size samples of random graphs of fixed
connectivity $K=3$ across the QSGT within the RITF model. Given the
high connectivity of these graphs, the applicability of the DMRG is
highly non-trivial. It is remarkable that in the case of non-random
couplings, we have found the number of retained states needed to
ensure a good accuracy to be very high, rendering the application of
the DMRG unfeasible.

The careful analysis of {\em pseudo-}critical points of finite size
samples constitutes a powerful tool to study a QSGT. As a recent
example, the work of Igl\'oi and coworkers \cite{Igloi_X07} studies
the distributions of, e.g., the average entanglement entropy and the
surface magnetization of finite size 1D samples. In this work we
provide measurements of, among other observables, spin-glass
susceptibilities, energy gaps, long distance spin-spin correlations
and entanglement entropies, and use them to obtain insight on the
mechanism of the transition. A full characterization of the QSGT,
nonetheless, is not attempted in this work, since it would require to
obtain statistics on a large number of disorder realizations.

The rest of the paper is organized as follows: Sec.~\ref{model:sec}
presents the model, and Sec.~\ref{dmrg} provides details on the
application of DMRG. Sec.~\ref{results} shows the numerical DMRG
results obtained. Sec.~\ref{conclusions} presents a discussion of the
results and some concluding remarks.

\section{Model} \label{model:sec} 

\begin{figure}
\epsfig{file=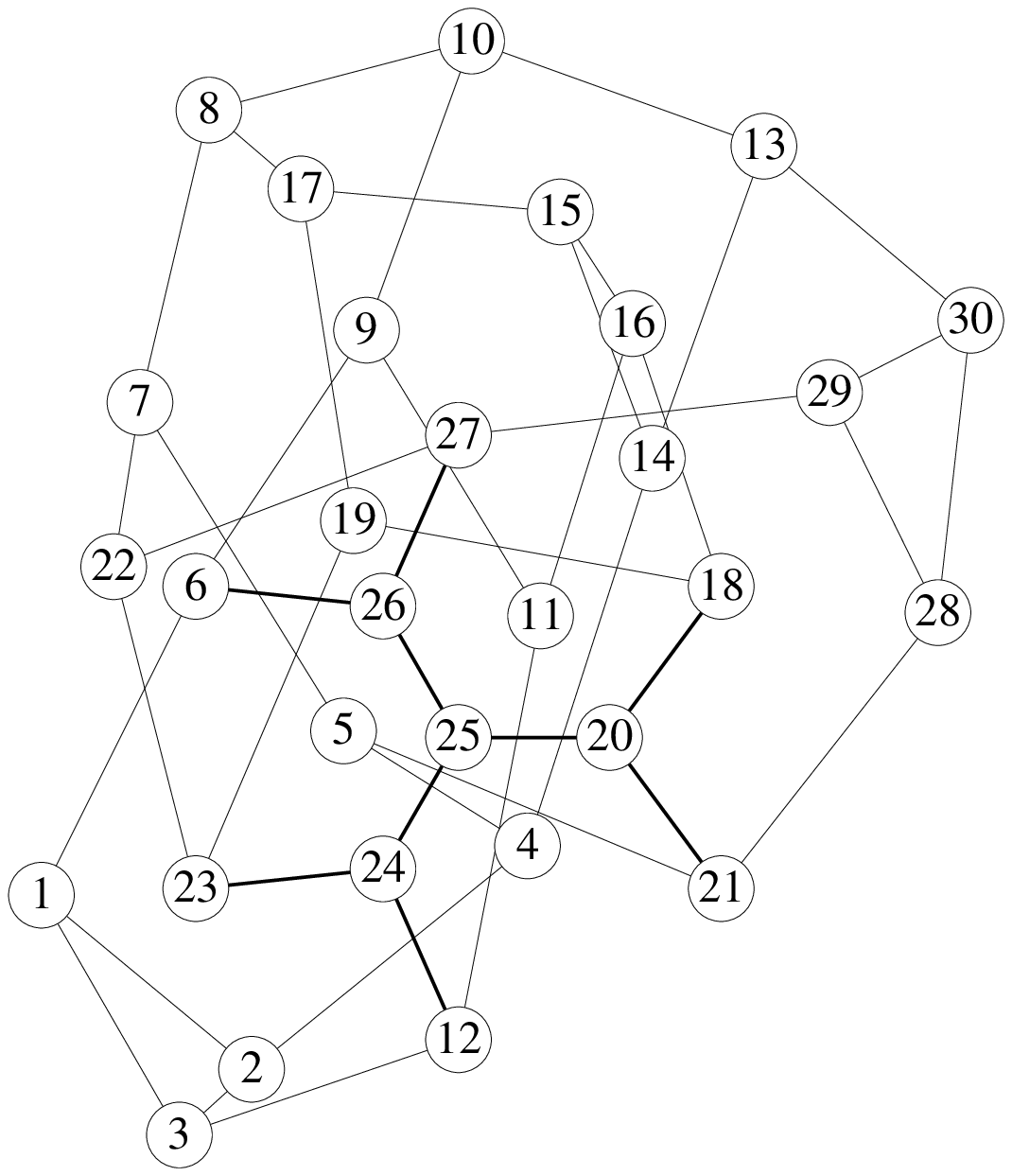,width=8.0cm,angle=0,clip=}
\epsfig{file=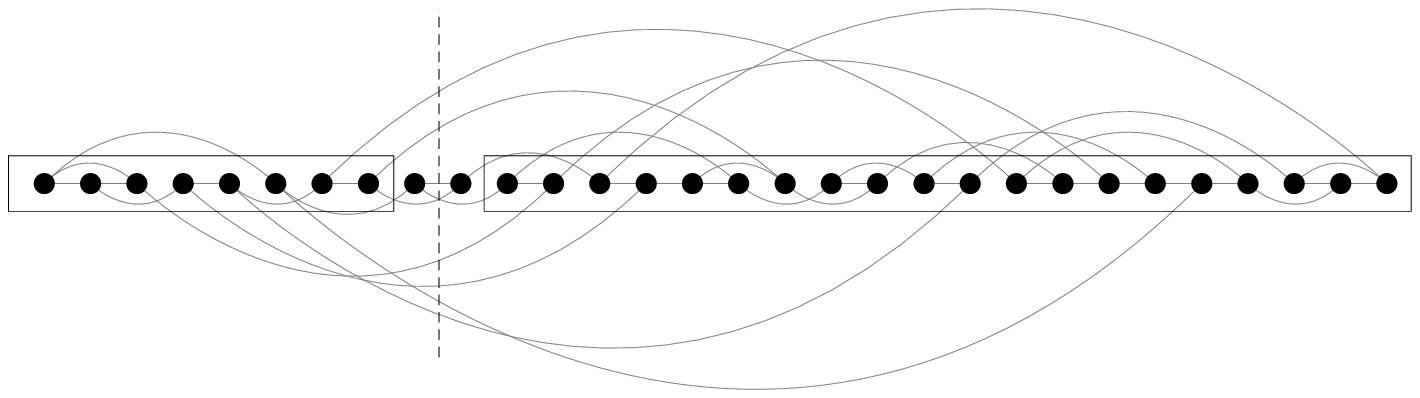,width=8.5cm,angle=0,clip=}
\caption{\label{graph:fig}
(a) A random graph with $N=30$ sites and connectivity $K=3$. Stronger
    bonds highlight the local Bethe lattice structure around site 25.
    Numerical labels denote the order in which they are visited
    along the DMRG path.  
(b) Deformation of the same graph to show the DMRG path as a straight
    line. The left and right blocks at a certain DMRG step are shown,
    and the dashed line cuts all the long-range links that would need
    reconstruction during that DMRG step, 9 in the example.}
\end{figure}

Let us consider the random Ising model in a transverse Field (RITF) on
a generic graph \cite{Chakrabarti_Dutta_Sen:book}:
\begin{equation} \label{itf}
H=-\sum_{\<i,j\>} J_{ij} \sigma^z_i \sigma^z_j - \sum_i h^z_i \sigma^z_i 
- \Gamma \sum_i \sigma^x_i  \;,
\end{equation}
where $\<i,j\>$ denotes pairs of neighboring sites $i$ and $j$ on the
graph. The values of $J_{ij}$ are uncorrelated random variables with
a uniform probability density distribution in the interval $[-1,1]$.
We will focus our study on randomly generated graphs, with $N$ sites
and a fixed connectivity $K=3$ \cite{Mezard_EPJ01,Marinari_JSTAT04},
i.e., each site $i$ is connected to $K=3$ other (randomly chosen)
sites $j$. An example of such a graph, with $N=30$ sites, is shown in
Fig.~\ref{graph:fig}.

It should be noticed that, locally and for large $N$, such graphs
resemble {\em Bethe lattices}, as we have highlighted with thicker
links in Fig.~\ref{graph:fig}. Nonetheless, the behaviour of finite
Bethe lattices is dominated by the boundary, unlike in our case, where
there is none. Moreover, trees have no loops and, therefore, present
no {\em frustration}, while random graphs with fixed connectivity do
contain loops and, therefore, have genuine frustration. It has been
proved, nonetheless, that short loops become rare as $N\to\infty$:
More precisely, the probability of finding a loop of any fixed size
$L$ falls to zero when $N\to\infty$ \cite{Marinari_JSTAT04}. The
physics of classical spin glasses in these graphs has been studied
using the cavity method by M\'ezard and Parisi \cite{Mezard_EPJ01}. 

Let $\epsilon_N$ be the absolute value of the disorder-averaged energy
per site of the classical ground state. Our numerical experiments show
that it increases from $\sim 0.6$ at $N=100$ up to $\sim 3/4$ for
larger $N$, which is the theoretical limiting value for $N\to\infty$
under the assumption that all links are satisfied.

We will take the product of eigenstates of $\sigma^z_i$, for all $i$,
as the {\em canonical basis} for our problem. In this basis, we denote
the state in which all spins point in the positive $x$-direction as
$\ket|\hat{x}>\equiv 2^{-N/2} \prod_i
(\ket|\uparrow>+\ket|\downarrow>)$.
Let us assume, for the moment, that $h^z_i=0$ for all $i$. For
$\Gamma=0$, all the non-diagonal terms in the Hamiltonian vanish.
Therefore, the ground state is given by the configurations with the
classical minimum energy. Disregarding accidental degeneracies, there
are two such configurations, related by a simultaneous flip of all
spins, $\sigma_i^z\to-\sigma_i^z$ for all $i$. The transverse field
$\Gamma$ may be considered as a {\em kinetic energy} coefficient,
providing a hopping term among the classical configurations. In the
$\Gamma\to\infty$ limit the transverse field term dominates and the
ground state tends to the state $\ket|\hat{x}>$, with all the spins
pointing in the positive $x$-direction, separated by a large gap
$\Delta\approx 2\Gamma$ from the first excited state. Let us remark
that all the components of the canonical basis have the same amplitude
in the state $\ket|\hat{x}>$.

By decreasing $\Gamma$ there is a certain value $\Gamma=\Gamma_c$ for
which the system undergoes a quantum spin-glass transition (QSGT). For
$\Gamma>\Gamma_c$, the system is in a quantum paramagnetic phase,
which presents no long-range order. Below $\Gamma_c$, the system is in
a quantum spin-glass phase, presenting a hidden long-range order which
may be detected by a number of observables. We will focus here on the
divergence of the {\em spin-glass susceptibility}:
\begin{equation} \label{xsg}
\chi_{SG}\equiv \frac{1}{N} \sum_{ij} 
\( \frac{\partial \<\sigma^z_i\>}{\partial h^z_j} \) ^2 \;,
\end{equation}
i.e., physically, a small longitudinal magnetic field $h^z_j$, applied
at site $j$, generates a magnetization response on each site $i$,
which is measured (and squared, so as to disregard its sign); the
results are summed over all sites $i$ and averaged over all sites $j$.
If the system is in a paramagnetic phase, the magnetization will be
proportional to $h^z_j$ and short-ranged in space, so that the sum
over $i$ and $j$ yields a finite value for $\chi_{SG}$. On the other
hand, on approaching the spin-glass phase, an infinitesimally small
longitudinal magnetic field $h^z_j$, localized at a single site $j$,
will eventually induce a finite magnetization over a long-range of
spins. This effect is at the origin of the divergence of $\chi_{SG}$.

We shall now discuss the numerical approach we have used to study this
system, and the results obtained.

\section{Application of the DMRG} \label{dmrg} 

The density-matrix renormalization group (DMRG) has proved to be an
accurate method to analyse the properties of 1D and quasi-1D systems
\cite{White_PRL92,Schollwock_RMP05,Martin_PRB02}. The DMRG may
be described as a variational method within the matrix-product states
(MPS), which constitute a low-dimensional subspace of the full Hilbert
space \cite{Rommer_PRB97}. A MPS may be expressed as
\begin{equation}\label{mps}
\ket|\Psi>=\sum_{s_1 \cdots s_N} 
{\rm Tr} \( A^{(s_1)} \cdots A^{(s_N)} \) \ket|s_1 \cdots s_N> \;,
\end{equation}
where each $A^{(s_i)}$ is a square matrix with dimension $m$, which
may be considered as the number of retained states per block when
splitting the system into a left and right parts. The total number of
variational parameters is less than $2Nm^2\ll 2^N$. The success of the
DMRG is related to the ability of these MPS to reproduce faithfully
the ground states of local 1D many-body problems for low values of $m$
\cite{Verstraete_PRB06}. If $m\to\infty$, any state of the Hilbert
space may be exactly represented as a MPS.

The minimum number of retained states $m$ is related to the
exponential of the von Neumann entanglement entropy of the DMRG block
\cite{Vidal_PRL02}. In a non-critical 1D system, this entropy is
bounded for all sizes, while it grows as $\ln L$ for a 1D critical
system of length $L$. It is believed that for a $D$-dimensional system
out of criticality, the entanglement entropy scales as $L^{D-1}$,
where $L$ is the shortest spatial dimension of the system
\cite{Sredniki_PRL93}. This estimate is known as the {\em area law}
and is believed to have logarithmic corrections at critical points.
An important practical consequence is that, in order to study a 2D
system, the number of retained states $m$ should grow as $\exp(L)$,
thus making the DMRG very inefficient.

Our system, on the other hand, is defined on a random graph of fixed
connectivity, $K=3$. We will show that this poses no problem to the
number of retained states $m$, which appears to remain manageable even
for $N=500$ as long as the couplings are random. However, the
implementation of the DMRG on such a model has required the following
technical refinements of the original method:
\begin{description}
\item[{\bf (a) Path selection.}] In a non-1D system, DMRG proceeds by
converting the system into an effective 1D problem with long-range
couplings. A path is chosen along the graph, which does not repeat
sites, and is considered to be appropriate if the number of broken
links between the left and the right blocks is kept low along a DMRG
sweep. Normally, the selection of a suitable path in a quasi-1D system
(e.g. ladders) is done by geometrical intuition. In our implementation
we have designed an automated procedure: a simulated annealing
algorithm is employed in order to minimize the number of broken
links. The full problem of finding the optimal path is computationally
very hard. Therefore, we do not aim at the exact optimum, but only to
a reasonably good local minimum. We have noticed that our
quasi-optimal path performs much better than a random path.
\item[{\bf (b) Perron-Frobenius criterion.}] The Hamiltonian of the
RITF on any graph fulfills, on the canonical basis, the conditions of
the Perron-Frobenius theorem, i.e., all off-diagonal components are
{\em non-positive}. Hence, all the ground state components must have
the same sign. Using the MPS representation of the ground state
obtained within the DMRG, it is always possible to reconstruct the
amplitude of any configuration $C=\{s_1 \cdots s_N\}$. The obtention
of the full $\ket|\Psi>$ is unfeasible, since it would require
reconstructing the amplitude of an exponentially large number of
configurations. Nonetheless, it is possible to pick up a few random
configurations $C$, and check that all their amplitudes have the same
sign. Whenever this criterion was not met ---a rather rare event---,
the calculation was repeated changing the random seed for the Lanczos
procedure on the first DMRG step. 
 
\item[{\bf (c) Wavefunction annealing.}] For large $\Gamma$ the ground
state $\ket|\hat{x}>$ is easily representable as a MPS, requiring a
single retained state. The DMRG works extremely well in this regime and,
therefore, our simulations are always started well within the
paramagnetic (large $\Gamma$) phase. The QSGT is approached by
repeatedly decreasing the value of $\Gamma$ by a small amount, always
using the previous ground state as a seed for the new calculation,
exploiting the wavefunction transformations suggested by White
\cite{White_PRL96}. This type of {\em annealing} of the ground state
provides a faster convergence and more accurate results for low
$\Gamma$.
 
\item[{\bf (d) Adaptive number of retained states.}] The number of
retained states $m$, and the number of DMRG sweeps $n_s$, are not
fixed in our algorithm. We set a maximum value for the sum of the
neglected eigenvalues of the density matrix in the RG truncation
($\eta\approx 10^{-6}$), and select $m$ accordingly. Moreover,
sometimes convergence takes more sweeps than usual ($n_s\approx
30-40$) in order to obtain machine precision in the convergence of the
ground state energy.

\item[{\bf (e) Energy gap measurements.}] When there is a symmetry in
a problem, e.g., under SU(2), it is usually possible to obtain the
first excited state as the ground state of a different {\em sector} of
the Hilbert space. In our case, lacking this, the best option has
proved to be the following one. For each DMRG step, after the ground
state $\ket|\Psi_0>$ had been found, it was ``promoted'' to a higher
energy by the following transformation of the Hamiltonian:

\begin{equation} \label{h.gap}
H\to H+\lambda\ket|\Psi_0>\bra<\Psi_0| \;,
\end{equation}

in such a way that the ground state of this new hamiltonian is the
former first excited state, as long as we take $\lambda>\Delta$, $\Delta$
being the gap we are looking for. The density matrix used for
truncation was built as a linear combination of the density matrices
for each state, with equal weights.
However, we should remember that only ground states of local
Hamiltonians are expected to be faithfully represented by MPS
\cite{Verstraete_PRB06}, and Eq.~(\ref{h.gap}) does not define a local
Hamiltonian. Therefore, the accuracy in the gap estimate is worse
than that in the ground state energy and in other observables.
\end{description}

\section{Results} \label{results} 

We have applied the modified DMRG technique to study a few samples of
random graphs with fixed connectivity $K=3$ and sizes ranging from
$N=50$ to $N=500$. 

A full characterization of the QSGT would require relevant statistics
on the disorder. Unfortunately, in order to obtain a high accuracy for
each sample, the required CPU-time is rather large. Therefore, we have
decided to focus on a precise study of a few realizations, in order to
gain insight on the mechanism of the transition for finite
samples. Most of our findings will be illustrated in the figures of
this section by showing in detail the results obtained for a sample
with $N=200$ sites. It should be remarked that the highlighted
features are typical of all the ensemble.

The numerical simulations were done with the DMRG algorithm explained
in section \ref{dmrg}, with a neglected probability tolerance
$\eta=10^{-6}$ and a tolerance on the convergence of the energy of one
part in $10^{10}$, for ten samples of each size ($N=$ 50, 100, 150,
200, 250, 300, 400, 500). Each sample is a different random graph,
always with fixed connectivity $K=3$, and the bonds $J_{ij}$ are
random and independent, uniformly distributed in the interval
$[-1,1]$.

\begin{figure}
\epsfig{file=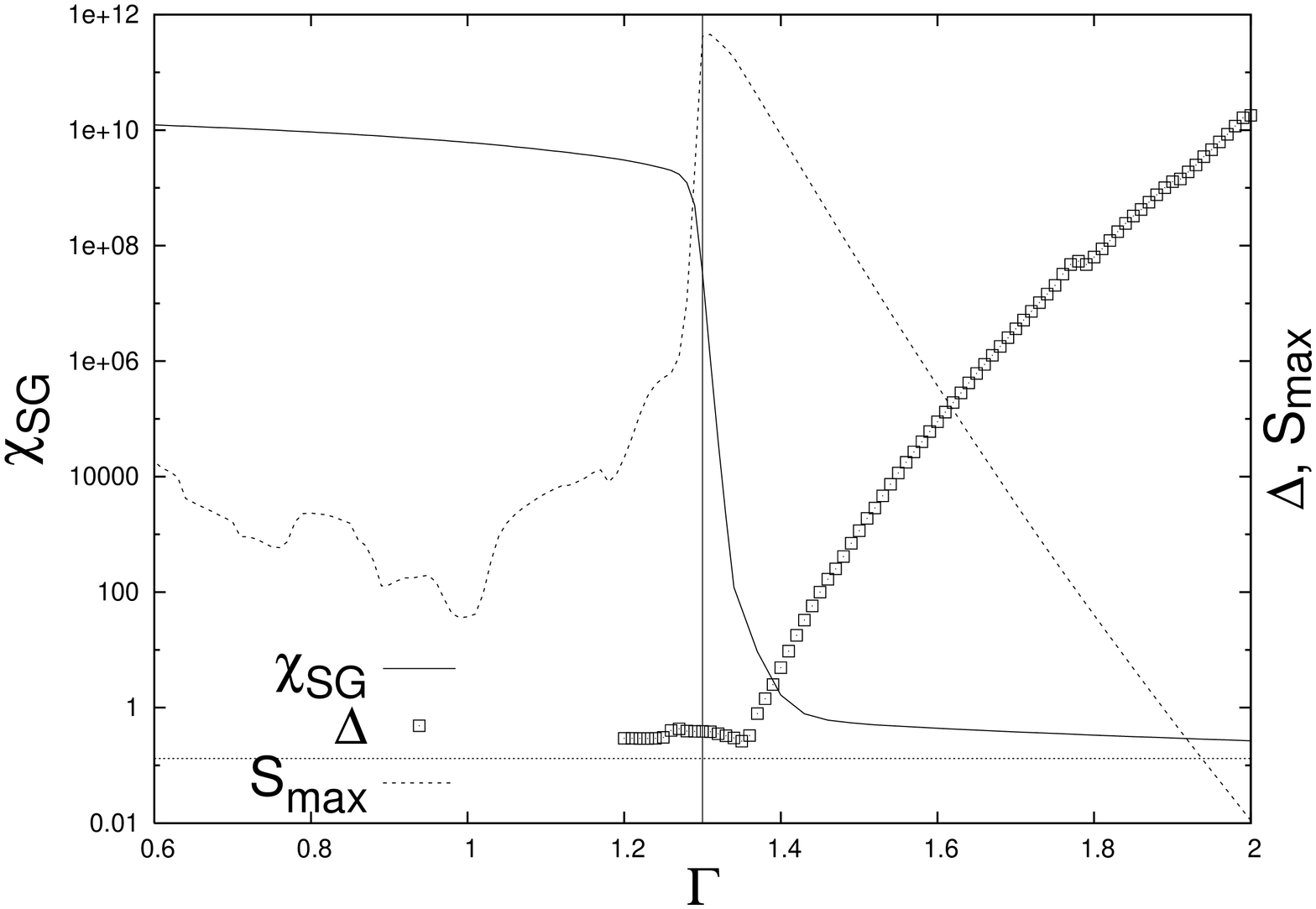,height=8.5cm,angle=270}
\epsfig{file=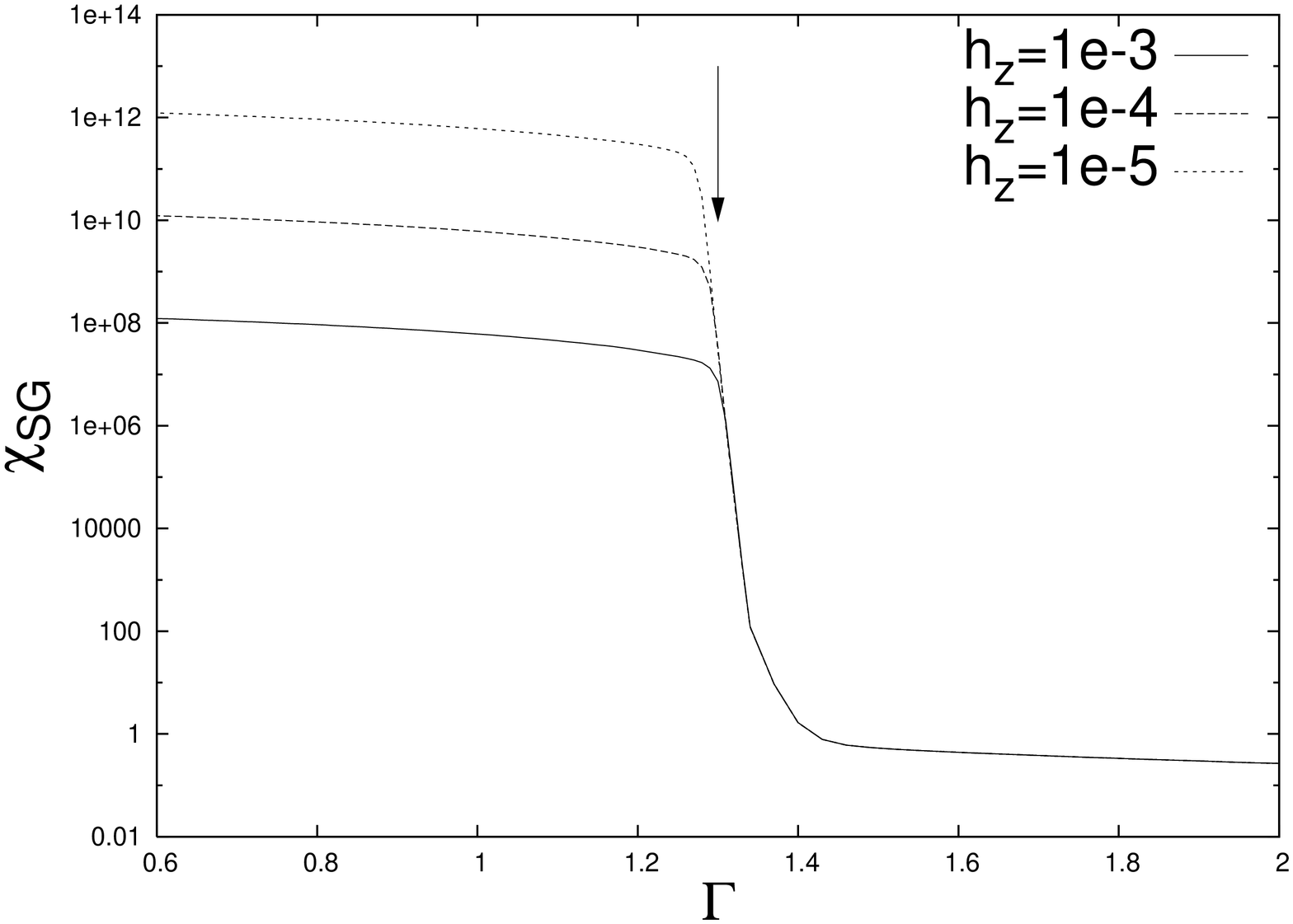,height=8.5cm,angle=270}
\epsfig{file=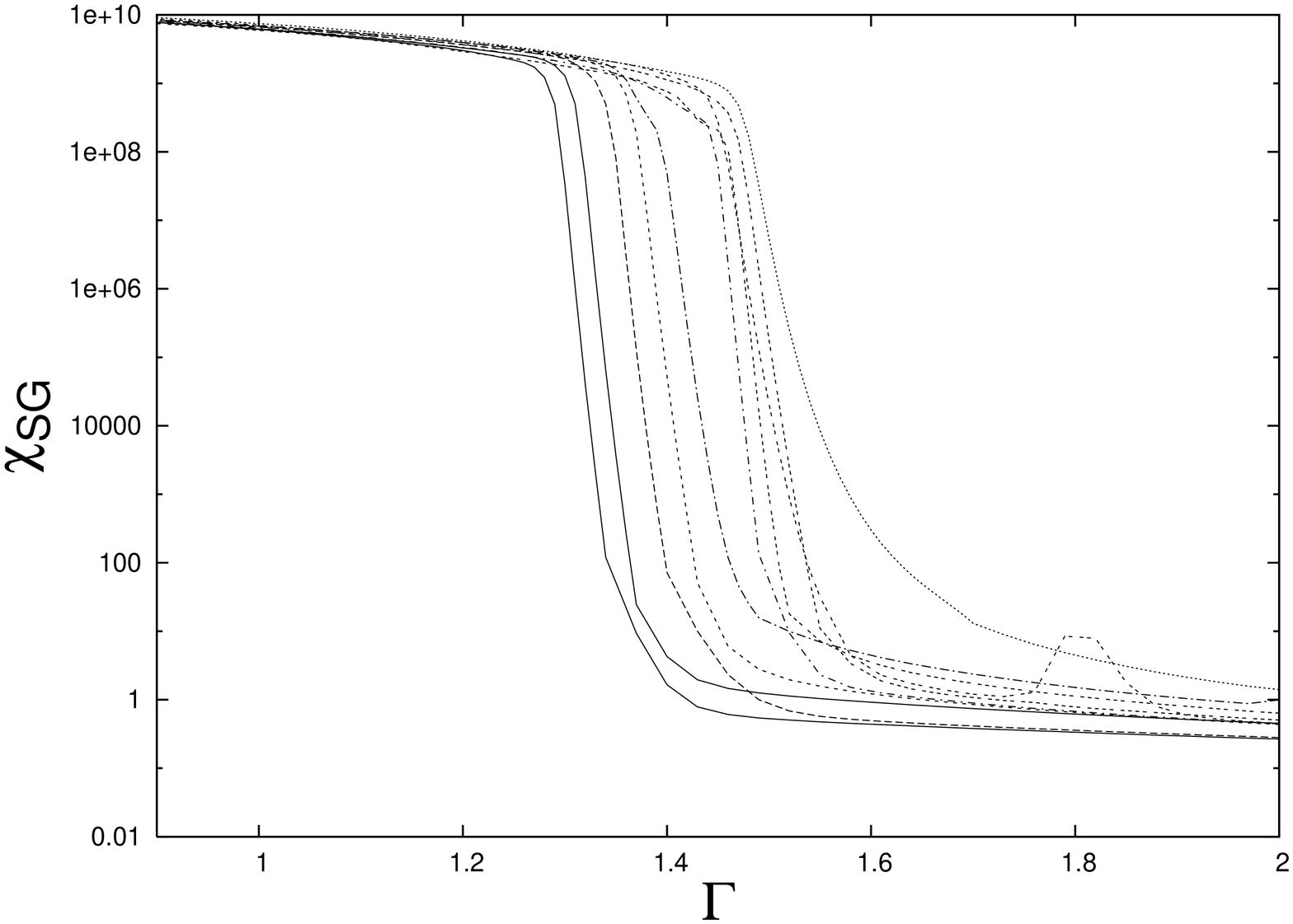,height=8.5cm,angle=270}
\caption{\label{xsg:fig} (a) For a random sample with $N=200$ sites,
the spin-glass susceptibility $\chi_{SG}$ is shown in logarithmic
scale, along with the maximum block entropy $S_{max}$ and the energy
gap $\Delta$, the two latest in arbitrary units. The divergence of
$\chi_{SG}$ is marked with the vertical line, which clearly coincides
with the maximum value of $S_{max}$. The energy gap comes very close
to zero (dotted line) precisely in that region.
(b) The fact that $\chi_{SG}$ saturates is related to the finite value
of the applied $h^z$ field. In this figure we can see how this
saturation value increases as $h^z$ is decreased.
(c) The divergence of $\chi_{SG}$ is shown for several samples, all
with $N=200$ sites.  }
\end{figure}

Figure \ref{xsg:fig}(a) shows three observables calculated for a given
instance of a $N=200$ graph, as a fuction of $\Gamma$. The most
relevant quantity is the spin-glass susceptibility, defined in
equation \ref{xsg}. A small magnetic field $h^z_{i_0}=10^{-4}$ is
applied at a single site $i_0$, and the magnetic response is measured
with the formula
\begin{equation} \label{xsg.i}
\chi^{(i_0)}_{SG} \equiv \sum_j \( \frac{\<\sigma^z_j\>}{h^z_{i_0}} \)^2 \;,
\end{equation}
which is found to increase very fast as $\Gamma$ approaches $1.28$
from above, and thereupon saturating at a very high value. Figure
\ref{xsg:fig}(b) proves the divergence of $\chi_{SG}$ by showing its
behavior at three different values of $h^z_{i_0}$
($h^z_{i_0}=10^{-3}$, $10^{-4}$ and $10^{-5}$): the saturation value
is seen to scale as $1/(h^z_{i_0})^2$. We estimate the {\em
pseudo-}critical value $\Gamma_c$ for the given sample as the value of
$\Gamma$ at which the slope of the susceptibility attains its
maximum. With this definition, $\Gamma_c$ appears to be almost
independent of the chosen site $i_0$, at our level of precision
$\Delta\Gamma=\pm 0.01$. Figure \ref{xsg:fig}(c), finally, shows
$\chi_{SG}$ for ten different samples with $N=200$ sites, differing in
the graph structure and in the choice of the couplings $J_{ij}$, and
diverging at different (sample dependent) values of $\Gamma_c$.

Figure \ref{xsg:fig}(a) also shows two other observables which point
towards the same value for $\Gamma_c$. The first is the block
entanglement entropy, defined as $S\equiv -{\rm Tr} (\rho
\log(\rho))$, where $\rho$ is the reduced density matrix for a part of
the system. It is measured for all the left-right decompositions along
a DMRG sweep, and its maximum value is denoted by $S_{max}$. This
value of $S_{max}$ is obviously dependent on the DMRG
path. Nonetheless, since this path has been chosen so as to minimize
the number of retained states, it is expected that it will also
minimize the maximum value of the entropy.
It should be emphasized that this observable attains its maximum
value, as a function of $\Gamma$, at the same $\Gamma_c$ which is
found by analyzing the spin-glass susceptibility. The relevance of the
entanglement entropy in order to characterize the critical point of a
QSGT has already been remarked in the literature \cite{Igloi_X07}. The
DMRG, being based on the obtention and analysis of the density matrix
of different parts of the system, is specially well suited for
measuring this observable \cite{DeChiara_JSTAT06}.

The other observable shown in figure \ref{xsg:fig}(a) is the energy
gap $\Delta$, which extrapolates to zero at a value of $\Gamma$ which
is close to $\Gamma_c$. The sign of the difference between these two
pseudo-critical points is sample dependent. This discrepancy is likely
to be a finite size effect.

\begin{figure}
\epsfig{file=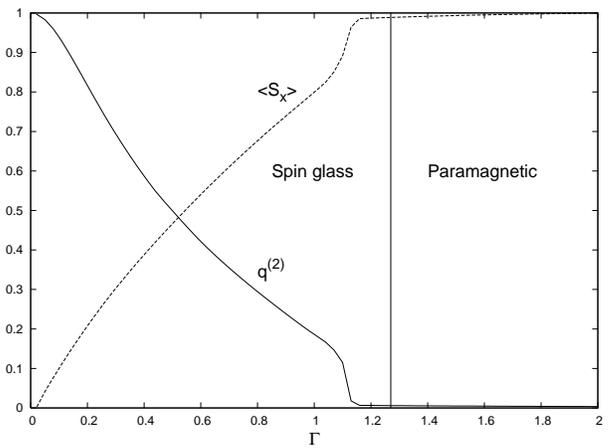,height=8.5cm,angle=270}
\caption{\label{szzsx:fig} The behavior of $\<S_x\>$ and the $q^{(2)}$
  order parameter is shown for the same sample of figure
  \ref{xsg:fig}. The two curves are strongly correlated, suggesting a
  transition at the same value of $\Gamma$. This value is
  substantially lower than the $\Gamma_c$ pointed by $\chi_{SG}$,
  and $S_{max}$, which is marked in the plot with a vertical
  line. }
\end{figure}

The next observable that we calculated for each sample is the average
value of $\<S^x\>$, see figure \ref{szzsx:fig}. For high $\Gamma$, the
transverse field is dominant, the ground state is close to
$\ket|\hat{x}>$ and the value of $\<S^x\>$ is close to 1. As we
decrease $\Gamma$, this value is reduced. It is remarkable that, at
the value of $\Gamma_c$ defined by $\chi_{SG}$, and confirmed by the
maximum block entropy, $\<S^x\>$ is never lower than $0.95$, at
variance with the 1D RITF, where it is normal to have values below
$0.5$.
%
%
Also in figure \ref{szzsx:fig} we have plotted the behavior of the
quantum equivalent of the $q^{(2)}$ order parameter, which measures
the magnitude of the long-distance spatial correlations
\cite{Binder_RMP86},
\begin{equation} \label{qea}
q^{(2)}\equiv 
\frac{2}{N(N-1)} \sum_{i< j} \<\sigma^z_i\sigma^z_j\>^2 \;.
\end{equation}

Our definition differs slightly from that used commonly in the
literature on classical spin glasses. Instead of the long-distance
behavior, we measure the global average behavior. They are equivalent
in a system with a Bethe lattice topology because the number of
neighbours of a given site at distance $d$ scales as $\exp(d)$. Also
note that these two observables, $\<S^x\>$ and $q^{(2)}$ have the
advantage that they do not vanish trivially in the absence of external
longitudinal magnetic field.

This quantity is close to zero within the paramagnetic phase, and
takes a non-zero value inside the spin-glass phase. Figure
\ref{szzsx:fig} shows that the behaviours of $q^{(2)}$ and $\<S^x\>$
are highly correlated, both pointing to a pseudo-critical point which
is, in many instances, sensibly lower than the one obtained with
$\chi_{SG}$ and $S_{max}$. This systematic discrepancy is not
explainable in the framework of classical physics, since the
fluctuation-dissipation theorem leads to the proportionality of
$q^{(2)}$ and $\chi_{SG}$ \cite{Binder_RMP86}. In the quantum case,
$\chi_{SG}$ is not equivalent to $q^{(2)}$, but to

\begin{equation}\label{qfluctdis}
\chi_{SG}={1\over N} \sum_{ij} \left[ \int_0^\beta d\tau
\<\sigma^z_i \sigma^z_j(\tau) \> \right]^2
\end{equation}
where the integral is performed on imaginary time, from $\tau=0$ to
$\beta$. Therefore, in quantum spin-glasses, $\chi_{SG}$ contains a
contribution from time-correlations, while $q^{(2)}$ only measures
spatial ones. Their different divergence points might suggest that
long-distance correlations develop at a higher value of $\Gamma$ in
time than in space.

\begin{figure}
\epsfig{file=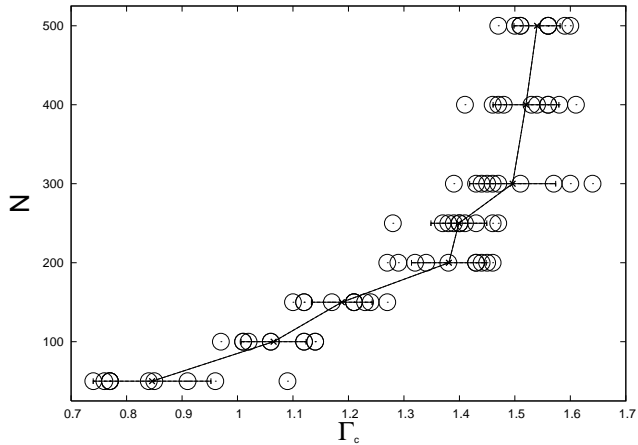,height=8.5cm,angle=270}
\caption{\label{gammac_vs_N:fig} The abscissa of each circle denotes
  the value of $\Gamma_c$ for a different sample, its ordinate being
  its size $N$. The crosses, joined by a line, denote the average
  values of $\Gamma_c(N)$ for each size. These average values seem to
  converge to a value around $1.5$.}
\end{figure}

At different sizes $N$, the behavior of the various samples is
qualitatively similar, the only difference being the sample-dependent
value of $\Gamma_c(N)$. It is possible to construct histograms
showing, for different $N$, the probability distribution of the
various $\Gamma_c(N)$. As a preliminary result, we mark in figure
\ref{gammac_vs_N:fig} by empty circles the $\Gamma_c$ values obtained
for 10 samples of different sizes: $N$=50, 100, 150, 200, 250, 300,
400, and 500. The crosses mark, for each size $N$, the average values
of $\Gamma_c(N)$, which apparently saturate at some point around
$1.5$.

\begin{figure}
\epsfig{file=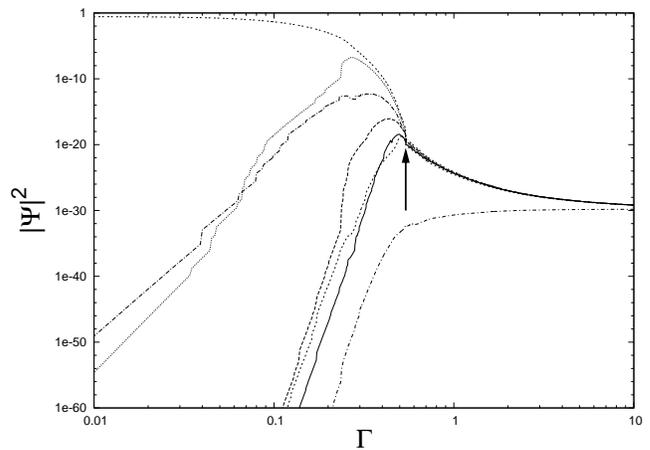,height=8.5cm,angle=270}
\caption{\label{wf:fig} Probabilities of various classical
  configurations within the ground state, as a function of $\Gamma$,
  for a sample with $N=100$ sites. Notice the log scale. The QSGT as
  obtained from the divergence of the spin-glass susceptibility is
  marked with an arrow. One of configurations is the one which
  minimizes globally the classical energy (for $\Gamma=0$). Another
  one is chosen at random. The rest are configurations with energies
  very close to that of the global minimum. It is noticeable how,
  after the transition, all the states but the random one increase
  their probabilities. Nonetheless, after some further decrease in
  $\Gamma$, all of them but the real minimum reach a maximum value and
  eventually fall to zero.}
\end{figure}

A useful clue to the physics of this system across the transition is
obtained by monitoring selected wavefunction components of the ground
state as a function of $\Gamma$, which can be easily done with the
DMRG. We illustrate this in figure \ref{wf:fig}, for a sample with
$N=100$ whose $\Gamma_c$ is marked by the arrow. One of the monitored
configurations is the classical ground state of the
system. (Naturally, there are two of them, related by a global
spin-flip, which we denote by $\phi^+$ and $\phi^-$. They are obtained
by measuring the values of $\<S^z_j\>$ at each site for very low
$\Gamma$ under the presence of a very small longitudinal field $h^z$
that splits the degeneracy between them. We will consider its weight
to be the sum of their probabilities.) The weight of the classical
ground state grows up to $1$ as $\Gamma$ is decreased. A second
monitored configuration is picked at random (dashed lower line in
figure \ref{wf:fig}): the weight of such a random configuration is
similar to that of all the others only for very large values $\Gamma$,
while it is definitely much smaller when $\Gamma$ approaches the
pseudo-critical point. The remaining monitored configurations in figure
\ref{wf:fig} are obtained by classical simulated annealing, i.e., they
are local minima of the classical energy. These configurations maintain a
high weight (similar to that of the optimal configuration)
across the transition, up to a value of $\Gamma$ below which their
weight decline markedly.
%
%

Therefore, the number of relevant classical configurations
contributing to the ground state changes drastically across
$\Gamma_c$. Deep into the paramagnetic phase, all of them have the
same weight, coherently bound within the $\ket|\hat{x}>$ state. This
state retains a high weight at $\Gamma_c$, as shown by the high values
taken by $\<S^x\>$ at that moment. Within the spin-glass phase, the
weight of the different configurations is gradually redistributed
according to their classical energies, until eventually only the
classical ground state remains.

It should be remarked that the DMRG was applicable in practice only in
the case of random couplings. For random graphs with fixed values of
$J$ (either ferro or antiferromagnetic), the number of retained states
needed for a similar accuracy increased in an order of magnitude,
rendering impractical the calculations.

\section{Discussion and conclusions} \label{conclusions} 
%

It should be remarked that a quantum phase transition may be observed
exactly in the MPS formalism with as few as two retained states per
block \cite{Wolf_X05}. The low values of $S_{max}$ measured in our
system points to the fact that the number of states involved in the
QSGT is reduced. Thus, in an attempt to explain the transition we
might build a simple-minded Ansatz consistent of a linear combination
of only two states: the classical ground state
$\ket|\phi>=\frac{1}{\sqrt{2}} (\ket|\phi^+>+\ket|\phi^->)$ and a {\em
background state} containing the rest of classical configurations:

\begin{equation} \label{background.states}
\ket|B>= \frac{1}{\sqrt{2^{N}-2}} \sum_{\phi\in S - \phi^{\pm}} \ket|\phi> \;.
\end{equation}

Neglecting matrix elements which are exponentially small for large
$N$, there is a very sharp crossover at $\Gamma_c(N)=\epsilon_N$, from
state $\ket|B>$ to state $\ket|\Psi>$. This last state clearly
presents a divergent spin-glass susceptibility. The physics of our
model is actually more complicated than this simple Ansatz, as shown
by the numerical estimate, $\Gamma_c(N)\approx 2\epsilon_N$ for large
$N$. 

An interesting question is how our system differs from the 1D RITF
analyzed by Fisher and coworkers \cite{Fisher_PRB95,Fisher_PRB98}. In
1D, duality arguments and a detailed RG analysis give a value of
$\Gamma_c$ satisfying $\log\Gamma_c=[\log J_{ij}]_{av}$, where
$[\cdot]_{av}$ indicates a disorder average. This equation leads to
values of $\Gamma_c$ much lower than those measured in our random
graph case (e.g., for our distribution of $J_{ij}$, Fisher's model
yields $\Gamma_c=e^{-1}$, while in our case it reaches
$\Gamma_c\approx 1.5$ for $N=500$ sites). The higher connectivity
seems to make a large difference in that respect. Moreover, in our
case, the value of $\<S^x\>$ at the transition is fairly high, about
$0.95$, while it is almost always below $0.5$ for RITF chains with the
same sizes. This means that the state $\ket|\hat{x}>$ is still
dominant in the paramagnetic phase at the moment of the divergence of
the spin-glass susceptibility.

In conclusion, we have extended the DMRG to make it suitable to study
the QSGT on random graphs. The main technical innovations are the
path-selection, which reduces the maximum number of retained states,
and the wavefunction annealing, which allows to use the ground state
for a certain value of $\Gamma$ as a seed to obtain the ground state
for a lower value. 

This modified DMRG algorithm has been applied to the measurement of
the energy gap $\Delta$, spin-glass susceptibility $\chi_{SG}$,
entanglement entropy $S_{max}$, long-distance spatial correlations
$q^{(2)}$ and average $x$-magnetization $\<S^x\>$ to high precision on
a few samples of different sizes, ranging from $N=50$ to $N=500$
sites. Remarkably, $S_{max}$ attains its maximum at the same value of
$\Gamma=\Gamma_c$ at which $\chi_{SG}$ diverges. This led us to
consider this value as our candidate for the pseudo-critical
point. The energy gap $\Delta$ vanishes in the surroundings of that
value for all realizations. In most disorder realizations, $q^{(2)}$
and $\<S^x\>$ start their crossover at a value of $\Gamma$ which is
inferior to $\Gamma_c$. This may suggest that long-range temporal
correlations develop at a higher value of $\Gamma$ than purely spatial
ones.

We would like to emphasize that our definition of $S_{max}$
constitutes an alternative approach to the entanglement entropy on
random graphs. Its relation to the standard block entropy, defined as
the entanglement entropy obtained by tracing out a site and its
neighbours up to a certain distance, remains as an open question.

We have also monitored the ground state probabilities of several
classical low-energy configurations, showing that, as $\Gamma$ is
reduced below $\Gamma_c$, these probabilities increase exponentially
until they attain a maximum value and then fall to zero, leaving the
classical minimum energy configuration as the only ground state
component as $\Gamma\to 0$.

The fact that the model on such disordered graph was amenable to
analysis within the DMRG is non-trivial. Both $S_{max}$ and the
maximum number of retained states per block increase slowly with the
system size. In a restricted sense, the system behaves similarly to a
1D chain, despite the high connectivity of the underlying graph. This
is a rather remarkable effect due to the disorder, which perhaps leads
to the selection of an effective 1D path of strong bonds. If the
couplings $J$ are not random, we have found that the number of
retained states per block increases in a much more pronounced way,
rendering the numerical DMRG approach impractical. It will be
interesting ---and is left to a future study--- to understand
quantitatively how the entanglement entropy at the transition behaves
as a function of the system size, both for random and non-random
couplings \cite{Refael_PRL04,Laflorencie_PRB05,Santachiara_X06}.

\begin{acknowledgments}
The author acknowledges G.E.~Santoro, R.~Fazio, R.~Zecchina, and
S.~Franz for instructive discussions.
\end{acknowledgments}



\end{document}